# Oxygen as a dual function regulator in MoS$_2$ CVD synthesis: enhancing precursor evaporation while modulating reaction kinetics

Keerthana S Kumar[1], Abhijit Gogoi[2#], Madhavan DK Nampoothiri[2#], Bhavesh Kumar Acharya[1], Manvi Verma[1], Ananth Govind Rajan[2*], Akshay Singh[1*]

[1]Department of Physics, Indian Institute of Science, Bengaluru, Karnataka 560012, India

[2]Department of Chemical Engineering, Indian Institute of Science, Bengaluru, Karnataka 560012, India

*Corresponding author: ananthgr@iisc.ac.in, aksy@iisc.ac.in

#These authors contributed equally.

## Abstract

Molybdenum disulfide (MoS$_2$) is a promising 2D transition metal dichalcogenide (TMD) for optoelectronics and quantum technology applications, but faces challenges in scalable synthesis and defect engineering. Oxygen-assisted chemical vapor deposition (O-CVD), which introduces in-situ oxygen during growth, shows excellent potential in resolving both issues at once. Although co-flowing oxygen shows improvement in growth, the underlying mechanistic role of oxygen remains unclear. In this work, a combination of oxygen dosing experiments, density functional theory (DFT) calculations, computational fluid dynamics (CFD) simulations, and ab initio molecular dynamics (AIMD) simulations, uncover the dual role of oxygen in O-CVD. Firstly, AIMD reveals that oxygen increases MoO$_3$ sublimation and enhances Mo$_3$O$_9$ supply. Concomitantly, DFT reveals that sulphur oxides, due to their bulkier nature than pure S$_2$, limit the formation of reactive MoS$_6$ intermediates. Subsequently, by experimentally varying the oxygen flow-interval, flow-rate, and flow-time, and correlating them with CFD simulations, we decouple oxygen's roles in source-poisoning prevention (i.e. MoO$_3$ evaporation) and growth regulation. We find that maintaining a low sulphur-to-oxygen (S:O$_2$) ratio at the MoO$_3$ boat and substrate during nucleation, and a high S:O$_2$ ratio at the substrate during growth is the key to obtaining large-area high-quality monolayer $MoS_2$, confirmed by our optical measurements. Based on our understanding, we present a kinetic phase diagram for $MoS_2$ synthesis, which will enable controlled oxygen dosing as a tuning parameter for scalable, defect-controlled monolayer $MoS_2$ synthesis.

**Main text**

Transition metal dichalcogenides (TMDs) are layered van der Waals materials with highly tuneable electronic and optical properties making them promising for next-generation optoelectronics and quantum technologies[1,2]. Among them, molybdenum disulfide ($MoS_2$) is extensively studied, owing to its direct bandgap, strong excitonic effects, and spin-valley physics enabling high-performance optoelectronic devices such as photodetectors[3], and ultra-scaled and flexible transistors[4,5]. Emerging work on defect engineering has enabled controlled tuning of optical properties and the realization of single-photon emitters for quantum photonics[6,7]. However, practical applications of these materials require overcoming two major hurdles: scalable synthesis of high-quality ML-TMDs, and efficient defect engineering[8]. Chemical vapor deposition (CVD) is a leading technique for scalable synthesis of ML-TMDs[9,10], but involves complex interplay of precursor chemistry and reaction kinetics, making reproducibility and precise control challenging[2]. Moreover, the high temperatures typically employed in CVD lead to high chalcogen defect density in as-grown samples. This has led to the need for a modified synthesis technique that can simultaneously provide high-quality scalable synthesis as well as defect control.

Oxygen-assisted chemical vapor deposition (O-CVD) is a modified version of CVD, where a controlled amount of oxygen is co-flown with the inert carrier gas, and has the potential to address the major challenges in conventional CVD[11–13]. An in-built advantage of O-CVD is oxygen's isoelectronic nature with sulphur, leading to oxygen passivation of vacancy-induced in-gap states in defective TMDs, particularly $MoS_2$. Oxygen chemisorption significantly enhances PL intensity[14] and improves electrical properties in $MoS_2$ without introducing additional deep defect states[15–17]. However, introducing a reactive species like oxygen in the CVD chamber can alter the reaction mechanism and kinetics of the process, which remain poorly understood. This lack of understanding of the reaction mechanism and reactor environment limits the efficient usage of in-situ oxygen as a tuning parameter for large-scale synthesis of high-quality ML-MoS$_2$. The reported roles of oxygen in O-CVD synthesis include prevention of MoO$_3$ precursor poisoning (sulphurisation of MoO$_3$) via re-oxidation, and reducing nucleation density by etching unstable nuclei[11,17]. In contrast, another study reports the beneficial impact of oxygen in stabilizing MoO$_3$ precursor on the substrate, and thus reducing the impact on the growth rate by desorption of precursor[10]. Notably, O-CVD uses a significantly lower MoO$_3$ precursor temperature (~ 530 °C)[12,14] c.f. conventional CVD (~ 750 °C)[18–20], an observation that remains unexplained. In other words, the mechanism by which oxygen interacts with the precursors and substrate, and the influence on the reaction mechanism, i.e., whether oxygen is a promoter or limiter, are unclear.

In this work, we explore the dual role of oxygen in MoS$_2$ O-CVD using complementary experimental and theoretical approaches. We employ a combination of time and dose-dependent O-CVD experiments, ab initio molecular dynamics (AIMD) simulations, density functional theory (DFT) calculations, and reactor-level computational fluid dynamics (CFD) simulations. Firstly, from AIMD simulations, oxygen was observed to enhance MoO$_3$ sublimation. At the same time, surprisingly, DFT calculations suggest that oxygen plays the role of a limiter by forming sulphur oxides that find it difficult to break stable MoO$_3$ trimers in the gas phase. Then, we measure flake size and nucleation density as a function of oxygen-dosing interval, concentration, and introduction time. Further, these oxygen-dosing experiments were simulated using CFD, and concentration profiles were directly correlated to the growth. The role of oxygen was analysed in terms of sulphur-to-oxygen (S:O$_2$) ratio at the MoO$_3$ boat and substrate, and a kinetic phase diagram identifying the different growth regimes is constructed. Optical spectroscopy studies were conducted to analyse the optical quality of the as-grown samples, and the importance of balanced use of oxygen in obtaining high-quality samples is emphasized. This study advances the understanding of oxygen's dual role in MoS$_2$ synthesis, beyond the simple consideration that oxygen can etch small MoS$_2$ nuclei, and encourages oxygen as a tuning parameter to achieve large-scale MoS$_2$ for optoelectronic applications.

**Results and Discussion**

**Role of in-situ oxygen in the synthesis of MoS₂**

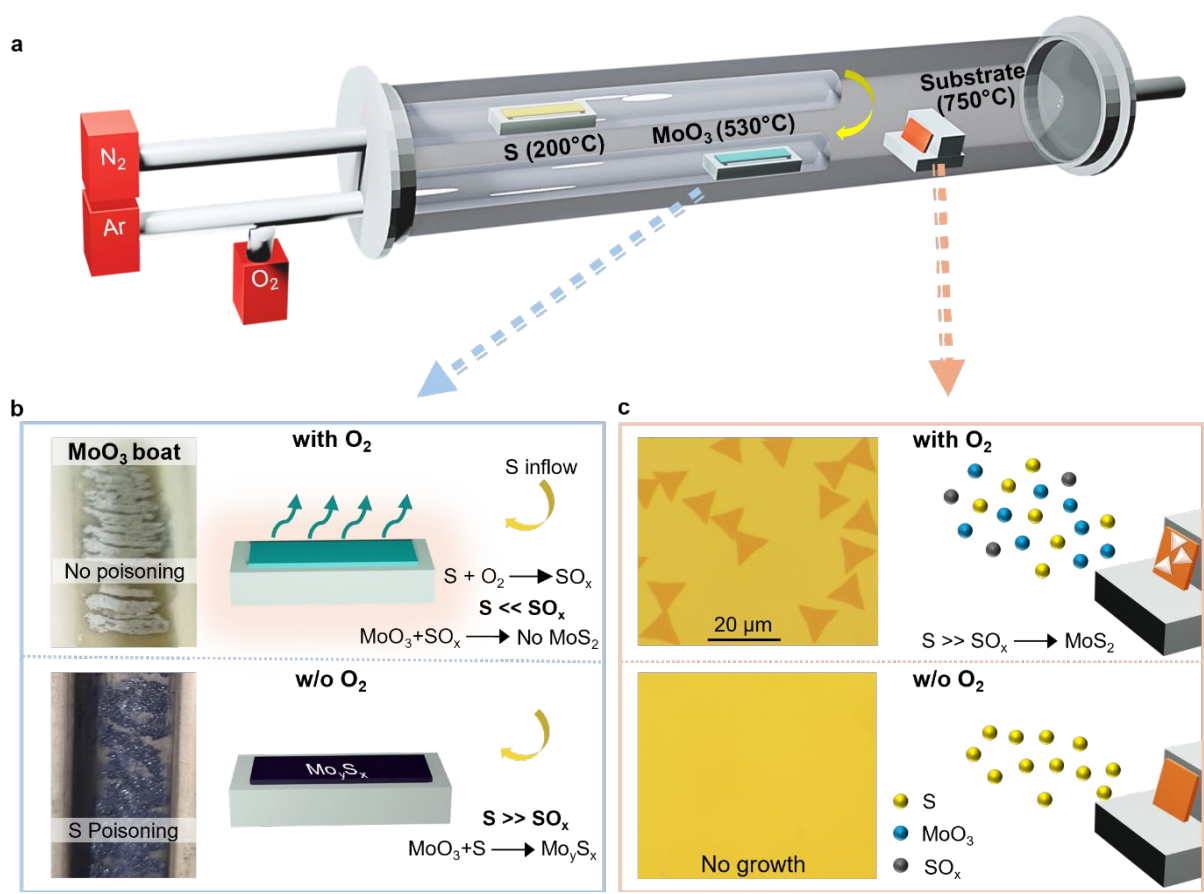

**Figure 1. Effect of oxygen in chemical vapor deposition (CVD) synthesis of MoS$_2$.** (a) CVD experimental set-up for multi-tube oxygen-assisted CVD (O-CVD) synthesis of MoS$_2$. Reactions occurring and species formed at the (b) MoO$_3$ boat and (c) substrate during CVD with and without oxygen.

To investigate the importance of oxygen in the O-CVD synthesis of ML-MoS$_2$, synthesis was performed under conditions of with and without oxygen. The process flow is shown in Supplementary Section S1. The CVD set-up has separate tubes for the MoO$_3$ and sulphur powders to avoid high sulphur flux at the MoO$_3$ source and thereby reduce precursor poisoning (Figure 1a, Supplementary Section S1, Methods). However, in the absence of oxygen, the MoO$_3$ powder in the boat was still poisoned (verified by the change in the colour from white to black), with no growth observed on the SiO$_2$/Si substrate (Figure 1b, c, Supplementary Section S1). This is reasoned to occur due to backflow of sulphur to the MoO$_3$ tube. Interestingly, when just 1 sccm of oxygen flow was used for first 35 mins (total process time = 30 (heating) + 10 (growth) mins) through the MoO$_3$ tube, the MoO$_3$ powder retained its original colour (white) and large ML flakes were observed on the substrate (Figure 1b, c). Notably, the as-grown flakes are of high-quality with ~ three times higher PL intensity than standard exfoliated samples, as reported in our previous work[14]. These high-quality ML-$MoS_2$ were reproducibly obtained at a reduced $MoO_3$ zone temperature, 530 °C in the presence of oxygen, as opposed to 750-800 °C in conventional CVD[18–20]. Precursor concentration follows an exponential relationship with temperature, making temperature a difficult and highly sensitive experimental parameter to optimise the growth. In contrast, oxygen concentration acts as a smoother parameter to tune the precursor evaporation at a lower temperature. Additionally, a high temperature in the precursor zone will influence the minimum achievable substrate temperature, since a large inter-zone temperature gradient can lead to heat leakage[21]. By enabling $MoO_3$ evaporation at reduced temperatures, oxygen allows the synthesis of $MoS_2$ on substrates with limited thermal stability.

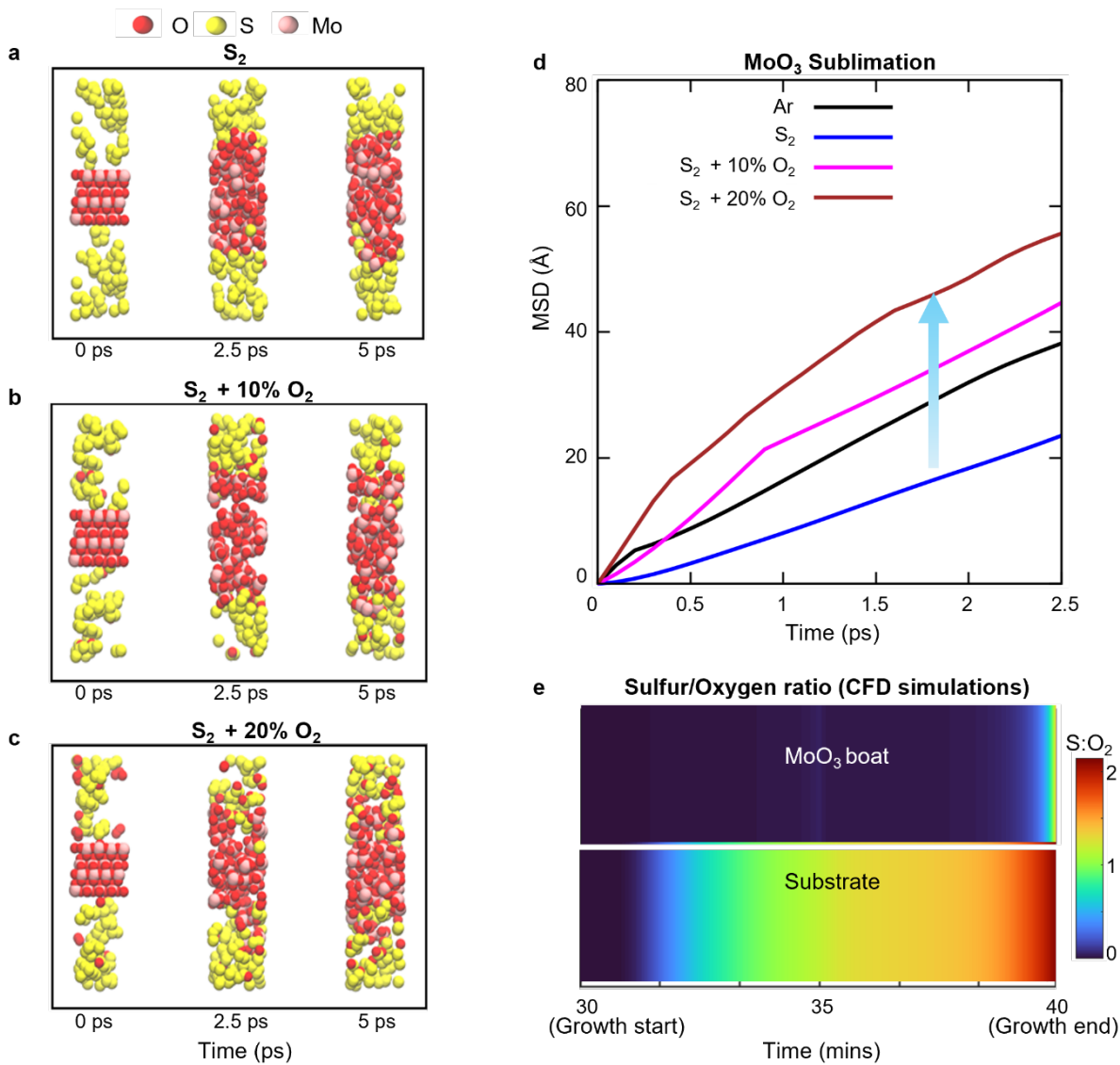

**Figure 2. Effect of the reaction environment on MoO$_3$ evaporation and dynamics of sulphur-to-oxygen (S:O$_2$) ratio.** Snapshots of the structures obtained from ab initio molecular dynamics (AIMD) simulations for the sublimation of MoO$_3$ sheets at 0, 2.5 and 5 ps in the presence of (a) only S$_2$, (b) S$_2$ + 10% O$_2$ and (c) S$_2$ + 20% O$_2$. Pink, red, and yellow spheres represent Mo, O, and S atoms, respectively. (d) Mean square displacement (MSD) of Mo atoms from each other in the presence of argon, sulphur and varied oxygen concentration with time. (e) Computational fluid dynamics (CFD) predictions of the S:O$_2$ ratio at the MoO$_3$ boat and substrate, at different points of time during the growth, with 35 mins of 1 sccm oxygen flow. The graph is plotted from 30 mins, i.e., from when sulphur starts evaporating.

Firstly, we examine why growth occurs at lower MoO$_3$ temperatures. We obtained insights into the sublimation rates of MoO$_3$ under varying oxygen concentrations, as well as in an argon environment using AIMD simulations (Methods). From the snapshots, the breakdown of the $MoO_3$ film is higher in the "$S_2$ + 10% $O_2$" (Figure 2b) and "$S_2$ + 20% $O_2$" (Figure 2c) cases compared to the "$S_2$" (Figure 2a) and "Ar" cases. This indicates a higher sublimation rate of $MoO_3$ film in the presence of molecular $O_2$ as compared to a pure sulphur or argon environment.. For a quantitative comparison, the mean-squared displacement (MSD) of the Mo atoms is computed during the course of the simulations (5 ps). The MSD ($\tau$) is computed by fixing an arbitrary time origin and a lag time $\tau$ as:

$$\text{MSD}(\tau) = \langle |r(\tau) - r(0)|^2 \rangle,$$

where $r(\tau)$ indicates the position of a particle at time $\tau$ and angular brackets indicate averaging over all pairs of Mo atoms. The slope of the MSD increases with oxygen concentration indicating significant structural deformations in $MoO_3$ (Figure 2d), thus quantifying the elevated sublimation rates in the presence of oxygen. This increased availability of precursor, in the presence of oxygen, explains how growth can be obtained at a lower $MoO_3$ temperature. Interestingly, the sublimation rate is higher in an argon environment as compared to a pure sulphur environment.

Next, to differentiate between reaction conditions at the $MoO_3$ boat and substrate, the variation of the $S:O_2$ ratio at these positions were studied using CFD simulations (Methods). Cut planes were defined on top of the $MoO_3$ boat and substrate (Supplementary Section S2), and the integrated concentration ratios of oxygen and sulphur were extracted (Figure 2e). The time-axis starts from 30 mins, since the reactor is set to reach the required temperature at that time. As expected, the $S:O_2$ ratio at the $MoO_3$ boat is lower than that at the substrate. This is because sulphur only reaches the $MoO_3$ boat through backflow, whereas oxygen directly flows through the $MoO_3$ tube. At the substrate, there is a continuous availability of sulphur which is ~ ten times of the total oxygen, increasing the $S:O_2$ ratio. It is due to the same reason that when separate mini tubes are not used for precursors, the $MoO_3$ source gets poisoned (Supplementary Section S1). As schematically depicted in Figure 1b and c, if the amount of $SO_x$ is higher than S, the $MoS_2$ formation reaction slows down, as explained below using first-principles calculations. Thus, our theoretical insights support the experimental observations that a low $S:O_2$ ratio at the substrate is detrimental to $MoS_2$ formation, whereas a higher ratio favours ML growth.

**Reaction pathways for oxygen-rich synthesis of MoS$_2$**

To understand how oxygen influences the reactions and to define a possible reaction pathway, reaction energetics calculations for MoS$_2$ formation were performed with MoO$_3$ and sulphur as precursors, and in the presence of oxygen. From previous studies, the effective precursors in the gas phase are Mo$_3$O$_9$[22] (i.e., trimerized MoO$_3$), O$_2$, and S$_2$[23]. After running AIMD simulations to study the reaction of the trimer ring (Mo$_3$O$_9$) with oxygen and sulphur, we identified the intermediates and constructed the reaction pathway (Figure 3a). The pathway for MoS$_2$ formation with MoO$_3$ and sulphur (S$_2$), without considering oxygen, is reported in previous studies[22]. For comparison, we name it as the oxygen-free pathway and the ones with oxygen are named as oxygen-rich pathways.

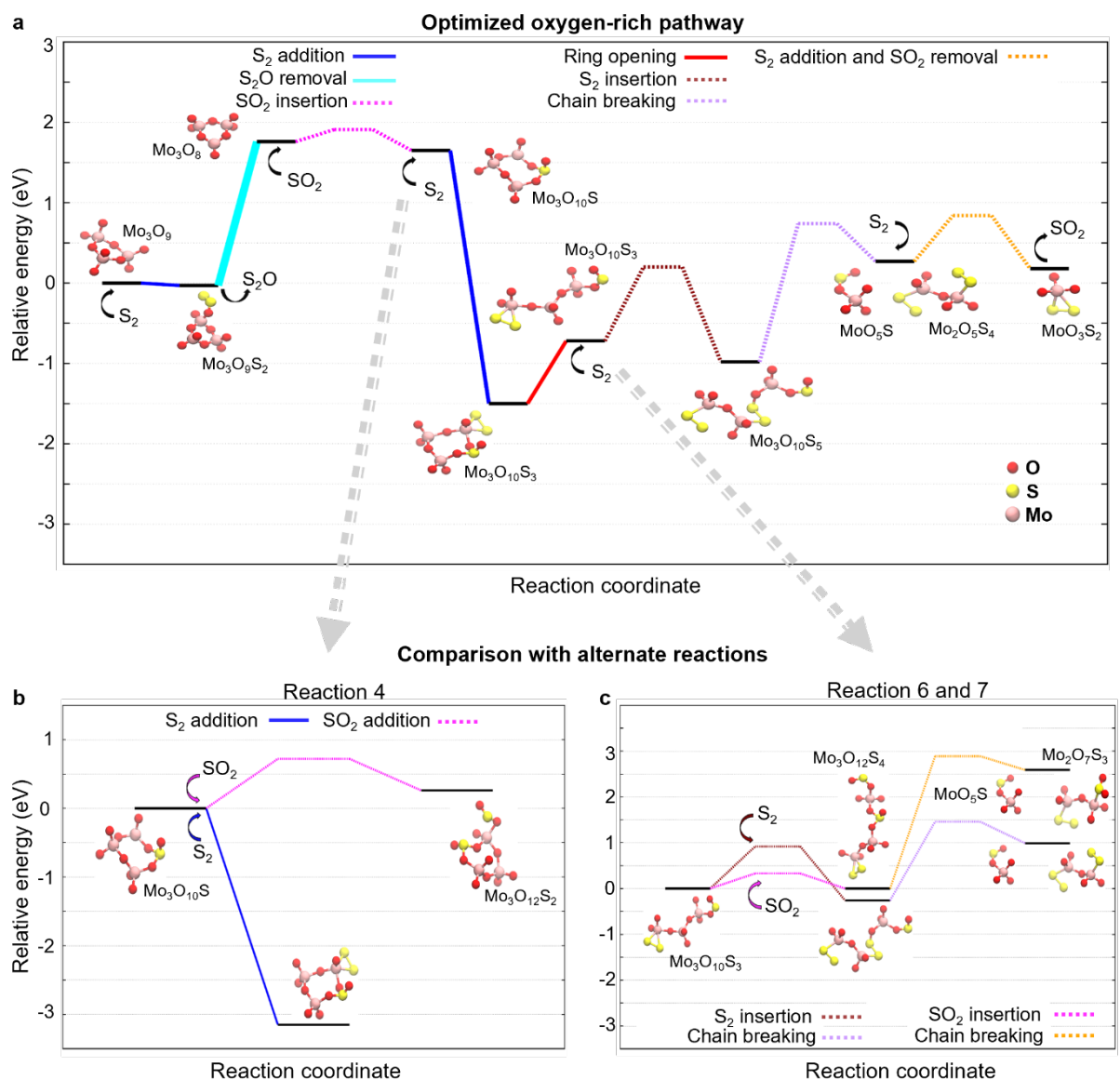

**Figure 3. Reaction pathway diagrams illustrating key steps in the formation of reactive precursor MoO$_3$S$_2$ from MoO$_3$O$_9$ in oxygen-rich environment.** (a) Optimized reaction pathway diagram of oxygen-rich pathway. Comparison of (b) reaction 4 and (c) reactions 6 and 7, with alternate reactions using sulphur oxides showing higher energy barriers. Dashed lines indicate reactions with transition states, while bold lines represent reactions without transition states, depicting uphill or downhill processes. A transition state is the highest energy point during a single reaction, where old bonds are being broken and new bonds are being formed simultaneously. It is a temporary, unstable arrangement of atoms that cannot be isolated like a stable molecule. Horizontal bold black lines denote intermediate species, with the adjacent smaller images showing the corresponding molecular structures. Pink, red, and yellow spheres represent Mo, O, and S atoms, respectively.

In the presence of oxygen, sulphur is prone to form various reactive species such as SO$_2$, S$_2$O and S$_2$O$_2$. Firstly, we investigated the possible sulphur oxides that can be present and reactive in the system. In the AIMD simulations for sublimation of MoO$_3$ sheets, we observed the formation of a significant number of SO$_2$ and S$_2$O$_2$ molecules. However, S$_2$O$_2$ molecule is very stable as well as larger in size, and thus would be unable to participate in the reactions.

We begin the reaction pathway simulations with S$_2$ addition into the Mo$_3$O$_9$ ring (reaction 1, Table S6, ; see below and Supplementary Section S5 for comparison with SO$_2$ addition). We explored combinations of different oxy-sulphide species to arrive at possible low-energy reaction pathways, enabling the simulation to comprehensively mimic the actual reactor environment. The optimized reaction pathway is shown in Figure 3a.

After the chain-breaking step (reaction 7), which is the rate-determining step, the oxygen-rich pathway proceeds with removal of SO$_2$ and addition of S$_2$ to MoO$_5$S, resulting in the formation of MoO$_3$S$_2$. The oxygen-free pathway reported in literature also leads to this intermediate after the chain-breaking step[22]. From there, the sulphurisation of MoO$_3$S$_2$ chain progresses through multiple S$_2$ insertions and S$_2$O removals, ultimately leading to the formation of MoS$_6$ (Supplementary Section S7) revealed as the final intermediate for MoS$_2$ growth[22].

Further, we re-investigated reactions 1, 4, 6, and 7 with SO$_2$ molecules (instead of S$_2$) to definitively understand the role of oxygen in the reaction. The direct insertion of SO$_2$ into the Mo$_3$O$_9$ ring (reaction

1) was observed to be energetically unfavorable (ΔE = 2.16 eV, Section S8). Similarly, for reaction 4 (Figure 3b, Table S8), $SO_2$ has a high energy barrier (0.72 eV) c.f. $S_2$ (no barrier). Next, $SO_2$ insertion into the chain in reaction 6 seems to have a lower energy barrier (0.33 eV), but leads to a very high energy barrier in the chain-breaking step (reaction 7), indicating the reaction will not proceed further (Figure 3c, Table S9). The resulting higher activation energy barriers c.f. reactions with sulphur, confirms the role of oxygen as a limiter in the reaction. We also note that oxygen can affect the sulphurisation process by reducing the partial pressure of sulphur through side reactions, and limiting the availability of $S_2$ for direct reactions with $Mo_3O_9$.

Thus, an oxygen-rich environment promotes $MoO_3$ evaporation, while a sulphur-rich environment is critical for $MoS_2$ formation. An oxygen-rich environment at the substrate during growth can oxidise sulphur, increasing the reaction energy barrier, and limiting growth. Hence, whether $MoS_2$ formation happens at either the precursor boat or the substrate depends on the dynamics of $S:O_2$ ratio throughout the growth. Further, understanding of this reaction mechanism also shines light on the timing of the nucleation process, which we discuss in the next section.

**Influence of oxygen in the reaction mechanism and kinetics of $MoS_2$ synthesis**

To probe the influence of oxygen on the growth kinetics and precursor ratio and to identify the nucleation window, we systematically studied the changes in flake size, shape, and nucleation density, as a function of oxygen flow-time, flow-interval, flow-rate, and total growth time [24–26].

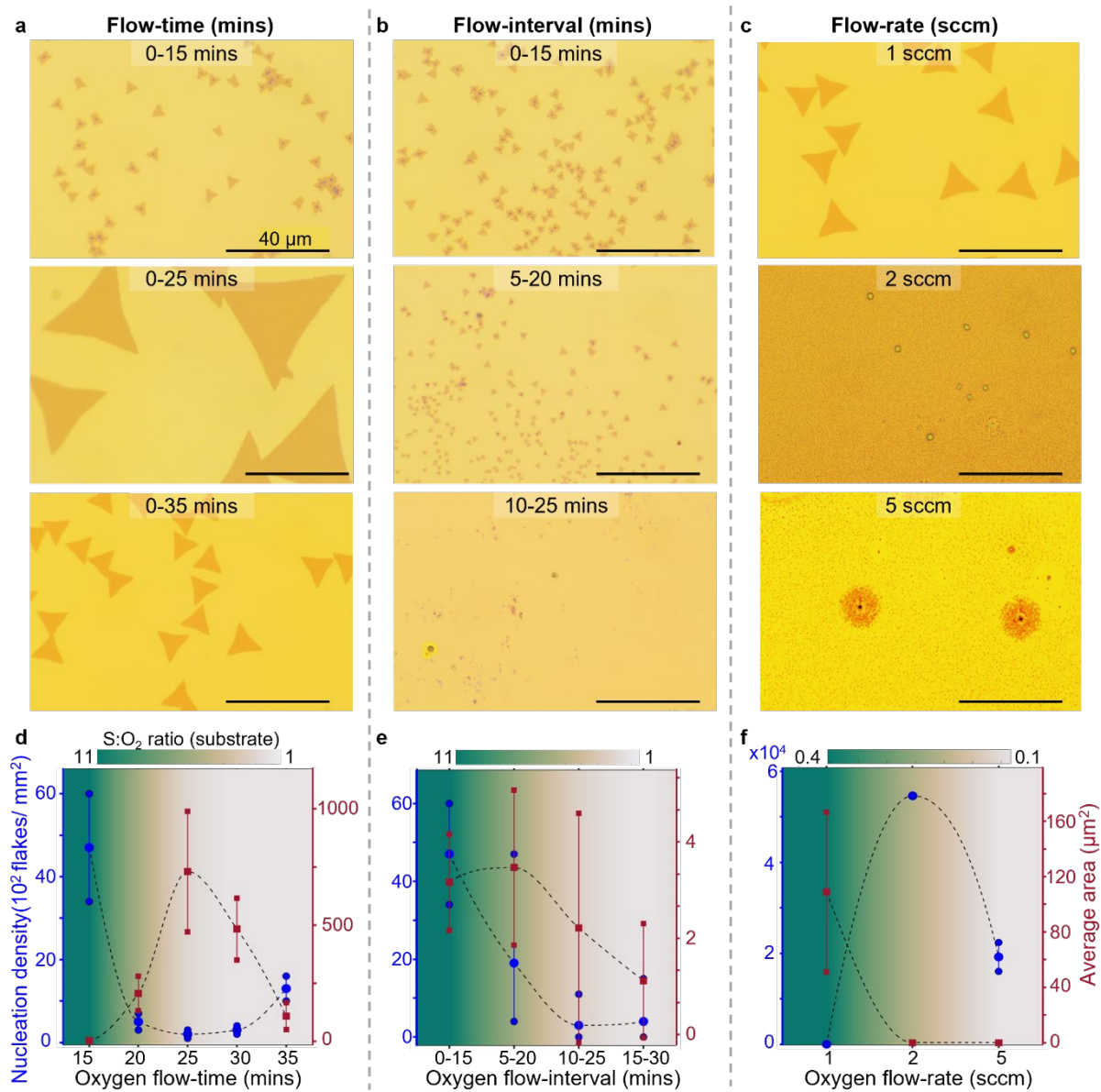

**Figure 4. Evolution of growth as a function of oxygen flow parameters.** Optical Microscopy (OM) images of substrate after growth with variable oxygen (a) flow-time, (b) flow-interval, and (c) flow-rate. Variation of nucleation density and average area with mean S:O$_2$ ratio at substrate, for variable oxygen (d) flow-time, (e) flow-interval and (f) flow-rate. Background colour in d-f indicates the S:O$_2$ ratio at the substrate. The ratio was calculated using CFD simulations and average concentration ratio over 1-3 mins of growth (approximate nucleation window) is shown. Circles and squares represent the average nucleation density and area respectively. The mean deviations of nucleation density, and average flake area (at similar positions on the vertically kept substrates) are represented by the circles and squares connected by vertical lines respectively. Sparse nucleation in some samples leads to skewed distributions, thus, plotting standard deviation leads to unphysical negative values. Hence, we truncate the lower bound to zero nucleation.

**a. Oxygen flow-time**

First, the total oxygen flow-time was varied from 0-15 till 0-40 mins (0 mins ~ beginning of heating). This experiment enables the observation of finer changes in growth, unlike exponential changes in the flow-rate study (Figure 4c). Initially, with increasing flow-time, nucleation density decreases, and size of flakes increases (Figure 4a). This is similar to earlier observed studies, where the etching effect of oxygen at higher concentration was reasoned to reduce nucleation density, leading to larger area growth[11]. However, at a flow-time of 35 mins, nucleation density increases, and the size of flakes decreases (Figure 4a), followed by etching at 40 mins (Supplementary Section S9).

CFD simulations find that the oxygen concentration at $MoO_3$ precursor boat and the substrate increases with increasing oxygen flow-time, correspondingly decreasing the $S:O_2$ ratio (Supplementary Section S9). Low $S:O_2$ ratio at the substrate during initial duration of growth (nucleation window) leads to reduction in nucleation density (Figure 4d). Later, when oxygen flow is stopped, the $S:O_2$ ratio increases at the substrate due to limited oxygen supply but continuous sulphur availability (Supplementary Section S9), facilitating the growth of ML-flakes. However, with a further increase in flow-time (35 mins), the nucleation density increases and flake size decreases, which can be attributed to increase in $MoO_3$ concentration and oxygen-limited growth respectively (Figure 4a,d). At 40 mins flow-time, the even higher concentration of oxygen oxidises the sulphur, and makes the growth rate lower than the etching rate, leading to etched flakes (Supplementary Section S9).

**b. Oxygen flow-interval**

To investigate space-dependent reaction dynamics, i.e. reactions taking place at precursor boat and substrate, we limited the oxygen flow to specific intervals, while keeping the total flow-time at 15 mins. The nucleation density monotonically decreases as the time interval is moved forward (Figure 4b, Supplementary Section S9). Further, as oxygen introduction is delayed even more, little to no growth was observed (Supplementary Section S9). The deviation from average values is higher in these samples due to non-uniform nucleation, present at only certain regions. Further, to probe the samples beyond the optical diffraction limit, scanning electron microscopy (SEM) was conducted and the observations were similar (Supplementary Section S10).

Unlike the flow-time experiment, the overall exposure of the $MoO_3$ boat and substrate to oxygen during these experiments stays the same, but is delayed (Supplementary Section S9). This delay in oxygen concentration at the $MoO_3$ boat increases the risk of precursor poisoning, affecting $MoO_3$ evaporation, and thus limits the growth. Hence, the presence of required amount of oxygen at $MoO_3$ precursor during evaporation is the most critical part in the O-CVD synthesis of $MoS_2$. Further, the nucleation density is directly proportional, and flake size is inversely proportional, to the amount of oxygen present at the substrate during nucleation, except for very high oxygen concentrations.

**c. Oxygen flow-rate and total growth time**

Finally, the flow-rate of oxygen was increased keeping the flow-time constant at 35 mins (maximum without etching). An increase in the flow-rate led to an exponential increase in nucleation density. This established that a higher oxygen concentration can lead to higher $MoO_3$ precursor availability and thus high nucleation density, despite having low $S:O_2$ ratio at the substrate (Supplementary Section S11). At the same time, the decrease in $S:O_2$ ratio results in significantly smaller flake sizes. There is a decrease in nucleation density at 5 sccm flow-rate possibly due to oxygen-induced etching (Figure 4c, f). Further, the central particles in as-grown flakes appear thicker at higher flow-rates, further emphasizing the role of oxygen in modulating the precursor concentration.

Interestingly, with the flow-rate and flow-time held constant, increased growth time led to increased flake size (Supplementary Section S11). This could be due to limited oxygen, leading to decreased oxygen concentration in the later stages of growth. Moreover, increasing the oxygen flow-rate at 20 mins growth time, caused some of the flakes to evolve into multi-edged stars indicating presence of excess precursor[24–26].

For the completion of our understanding, we repeated these growth experiments on sapphire substrates. We observed similar trends where an increase in oxygen flow-time led to reduced nucleation density and increased size of flakes, and increased oxygen flow-rate led to increased nucleation density (Supplementary Section S11). We note that size of flakes is larger in case of growth on sapphire substrate, c.f. on SiO$_2$/Si.

**Optical properties of as-grown MoS$_2$**

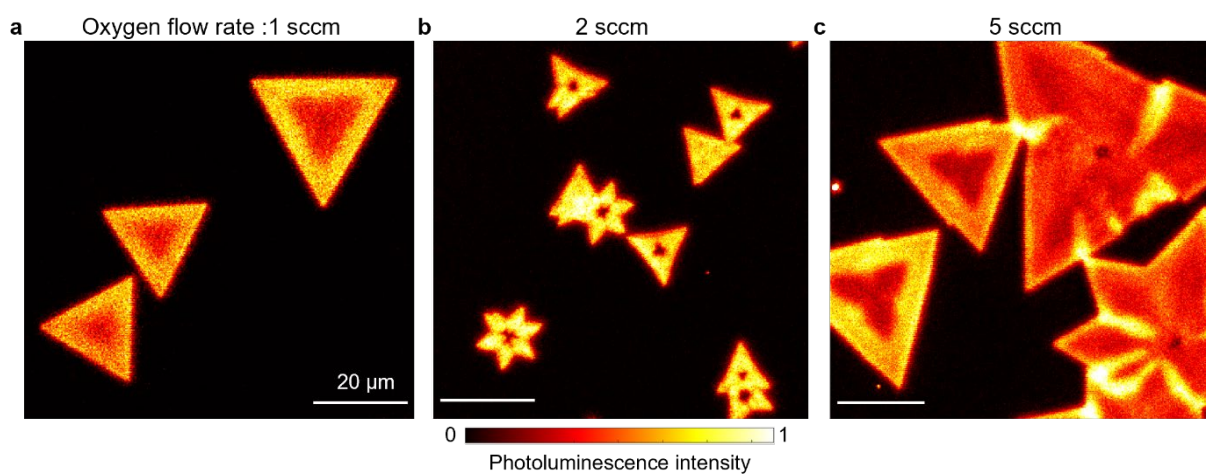

**Figure 5. Room temperature photoluminescence (PL) map of samples synthesized at different oxygen flow-rates.** Room temperature PL map of samples synthesized at an oxygen flow-rate of (a) 1 sccm, (b) 2 sccm, and (c) 5 sccm. Growth time is fixed at 20 mins for a-c.

Room temperature PL mapping was performed, to understand the synthesized samples beyond optical microscope images. The samples were synthesized under different oxygen flow-rates, and for 20 mins growth time to get larger flakes. We focus on PL variations across different flakes, and within a flake for a particular growth substrate. At low oxygen flow-rates (1-2 sccm), the PL intensity and spatial profile is largely similar for different flakes on the substrate (Figure 5a, b). In contrast, the 5 sccm sample shows noticeable flake-to-flake PL variation (Figure 5c). All samples have nearly similar room temperature PL spectra, with small peak shifts (Supplementary Section S12). Importantly, we observe that for both 1 sccm and 5 sccm samples, the edges appear to be brighter than the interior of the flakes. This indicates an inhomogeneous distribution of strain and defects in these samples[27,28]. Meanwhile, the 2 sccm sample showed relatively uniform PL intensity within the flake. The darker centres observed in some flakes are from thick central particles (Supplementary Section S11).

The comparison of PL maps indicates that both low and high oxygen flow-rate regimes are undesirable for optically high-quality growth. This observation re-iterates the role of oxygen in balancing nucleation, growth, etching, and defect control. Hence, by careful optimisation depending on the target application, oxygen can be used to tune the morphology and optoelectronic properties of ML-MoS$_2$.

**Kinetic phase diagram for MoS$_2$ growth**

So far, we have uncovered the complex interplay of sulphur, MoO$_3$, and oxygen in regulating the growth of MoS$_2$. Based on this nuanced understanding, we propose a kinetic phase diagram of the MoS$_2$ synthesis system (Figure 6). This encompasses the influence of oxygen on nucleation, growth, and morphology of ML-MoS$_2$, and is represented by S:O$_2$ ratio (horizontal axis), with the MoO$_3$ concentration on the vertical axis. Here, the ratios and concentrations are chosen to be the values at the substrate.

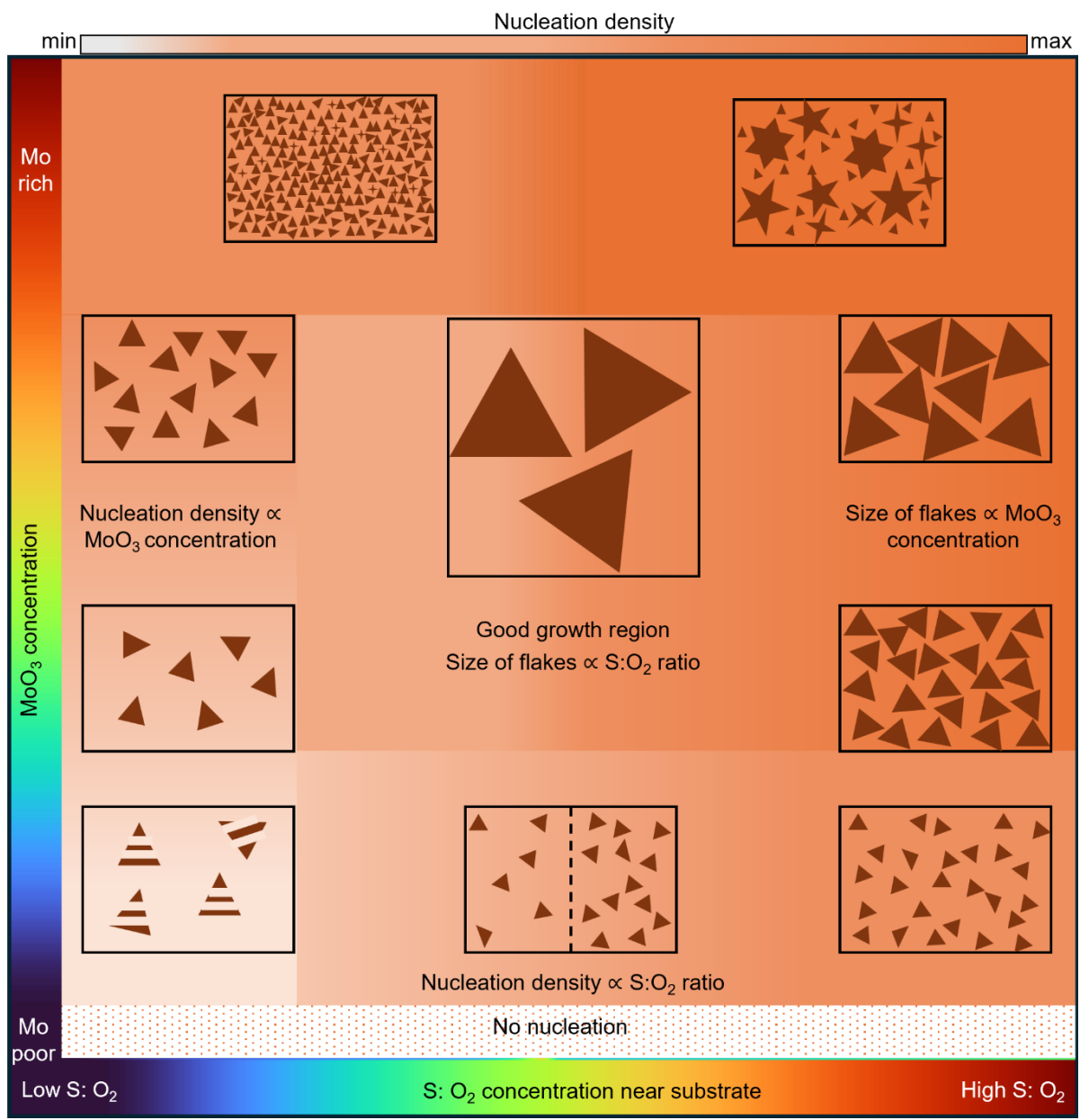

**Figure 6. Kinetic phase diagram of the influence of oxygen near precursor, in nucleation and in the growth and morphology of monolayer MoS₂.** The evolution of nucleation density and morphology with variation in S:O$_2$ ratio (horizontal axis) and MoO$_3$ concentration (vertical axis), evaluated at the substrate.

At a very low MoO$_3$ concentration, nucleation will not happen even at a high S:O$_2$ ratio. The presence of oxygen is critical at the MoO$_3$ boat to aid in evaporation, preventing poisoning, and thus ensuring nucleation. Then, at small MoO$_3$ concentrations, if the S:O$_2$ ratio is very low at the substrate, the growth rate will be lower than the etching rate, leading to etched flakes. As the MoO$_3$ concentration increases, the nucleation density will be dependent on it, with the flake size limited by the availability of sulphur. Similarly, for a particular MoO$_3$ concentration, the nucleation density will be determined by S:O$_2$ ratio at the substrate, with size limited by the MoO$_3$ availability. Interestingly, at very high MoO$_3$ concentration and considerably high S:O$_2$ ratio at the substrate, multi-edged flakes would be obtained due to unbalanced precursor ratio during the growth.

It is noteworthy that the concentration profile is dynamic throughout the growth. The key idea of the kinetic phase diagram is to provide a guide in tuning the concentration profile throughout the growth to ensure required nucleation density and flake morphology. For example, to obtain large area monolayers, the growth should be designed to get a low nucleation density (low S:O$_2$ ratio at substrate in the beginning of growth, low S:O$_2$ ratio at MoO$_3$ boat) and large size of flakes (high S:O$_2$ ratio at substrate at later times, low S:O$_2$ ratio at MoO$_3$ boat).

**Conclusions**

In summary, through a complementary combination of careful O-CVD experiments, detailed quantum mechanical theory (AIMD and DFT), and extensive transport simulations (CFD), we have clarified the mechanism of O-CVD synthesis of monolayer MoS$_2$. Our experiments show that having sufficient oxygen at the MoO$_3$ boat at the beginning of growth is critical to ensure nucleation. This is attributed to oxygen-induced enhancement in MoO$_3$ evaporation and prevention of precursor poisoning. AIMD simulations suggest increased MoO$_3$ sublimation in presence of oxygen, which also explains the lower MoO$_3$ evaporation temperature required in O-CVD c.f. conventional CVD, leading to a lower thermal budget requirement. On the other hand, oxygen leads to the production of sulphur oxides, which reduce the rate of formation of MoS$_2$ growth species, such as MoO$_3$S$_2$ and MoS$_6$. Thus, oxygen plays a dual yet complementary role by accelerating MoO$_3$ evaporation, while also being a limiter in the reaction mechanism.

During nucleation, a low S:O$_2$ ratio at the substrate leads to increased sulphur oxide formation over sulphur, limiting MoS$_2$ formation, thus decreasing the nucleation density. On the other hand, in the growth phase, high S:O$_2$ ratio is beneficial to obtain large-area monolayers. Thus, dynamic control of oxygen is critical to balance the contrasting needs of nucleation and growth.

Overall, understanding reaction pathways, variations in S:$O_2$ ratio, and $MoO_3$ evaporation rates in the presence of oxygen, enables in-situ tuning of precursor supply and reaction rates through minor oxygen parameter optimization. Controlling the precursor ratio by changing macro-parameters like temperature or precursor amounts is challenging and can cause precursor deficiency, negatively affecting growth. In contrast, careful control of in-situ oxygen helps in poisoning prevention and maintains $MoO_3$ evaporation without disrupting the growth. Thus, this work uncovers the paradoxical role of in-situ oxygen as a tuning parameter and provides a pathway for its effective use in scalable, high-quality $MoS_2$ synthesis.

**METHODS**

**Oxygen-assisted chemical vapor deposition (O-CVD) and microscopy.**

A three-zone furnace was used (Ants Ceramics), having three independent temperature zones of width 150 mm separated by insulation zones of width 75 mm. Two separate mini tubes (OD: 20mm, ID: 17 mm) were used for sulphur and $MoO_3$, extending from first zone till the centre of second zone (375 mm) inside the main quartz tube (OD: 60 mm, ID: 55 mm, length : 1500 mm). Sulphur was kept in the first zone at 200 °C, $MoO_3$ at 530 °C in the second zone, and the substrate (285 nm $SiO_2$/Si prime substrate) in the third zone at 750 °C. The substrate was kept vertically to ensure uniform precursor concentration along the height of the substrate[29]. 50 sccm argon was used as carrier gas for sulphur and a mixture of 50 sccm nitrogen and 1 sccm oxygen were flown through the $MoO_3$ tube. To minimise contamination, two cycles of evacuation and argon filling were conducted, followed by 10 mins of argon flushing, prior to start of the process. Each zone is heated to the required temperature in 30 mins and maintained at that temperature for 10 mins to obtain ML growth (except for experiments with variable growth times and flow-rates). The furnace is then opened for fast cooling in the presence of 200 sccm Argon. The images of MLs were taken using Olympus BX 51M optical microscope at 100X magnification. Scanning electron microscopy was performed using an Ultra55 FE-SEM Karl Zeiss microscope with 5 kV accelerating voltage.

**Multiphysics computational fluid dynamics (CFD) simulations**

Multiphysics simulations were performed using COMSOL Multiphysics. The physics inside the CVD tube was modelled using the CFD and Heat Transfer modules with the submodules laminar flow, transport of dilute species, transport of concentrated species, and heat transfer of fluids. Even though argon was used in the CVD experiments as the carrier gas in sulphur tube, it was replaced by nitrogen for ease of computation, with similar results expected. Thus, in the simulations, nitrogen was flown through the sulphur and $MoO_3$ tube, while oxygen was flown through the $MoO_3$ tube. A 'coarser' mesh with 20329 elements and an average quality of 0.8275 (ranges from 0-1) was used, which required a simulation time of ~ 4.7 hours. Then, cut planes were defined at the $MoO_3$ boat and substrate, and surface integration of the required species (oxygen, sulphur) was carried out to extract the concentration profiles.

**Ab-initio molecular dynamics (AIMD) simulations**

**a. Convergence study for the AIMD simulation parameters**

The model we employed was a cubic simulation box of size 20 Å consisting of 20 $O_2$, 5 $S_2$, and 1 $Mo_3O_9$ molecules for an oxygen-rich environment. Based on the DFT energy convergence study for the model simulating the oxygen-rich environment, a Gaussian smearing width of 0.01 eV and a Gamma point calculation with a basis set corresponding to a 520 eV kinetic energy cutoff (against a benchmark planewave cutoff of 550 eV) were found to be sufficient to reach the convergence criteria of around 1-2 meV/atom. These calculations are documented in Tables S1, S2, and S3. We used the Vienna ab initio Simulation Package (VASP.6.2.0)[30–32] implementation of DFT based on the projector augmented wave (PAW) method[33,34]. The Perdew–Burke–Ernzerhof (PBE) functional[35] was used to capture the electronic exchange-correlation interactions. Based on the convergence study summarized in Tables S1, S2, and S3 in the Supplementary Information, a gamma-point mesh was used for Brillouin zone sampling. The AIMD simulations were run at a temperature of 2500 K with a time step of 1 fs in the NVT ensemble using the Nosé-Hoover thermostat for temperature control[36]. The AIMD initial configurations were generated using Packmol[37]. The AIMD simulations were performed for 70 ps. Considering the computational expenses, spin polarization is not considered in the Nudged elastic band (NEB) and AIMD simulations.

**b. Calculation of MoO₃ sublimation in the presence of sulphur and oxygen**

The AIMD simulations were performed at 2500 K using the canonical (NVT) ensemble. The simulations were performed for 5 ps with a time step of 1 fs. Four different cases were considered, involving a MoO$_3$ (001) film (Mo$_{64}$O$_{192}$) immersed in: (1) sulphur molecules ("S$_2$" case); (2) a mixture of sulphur and oxygen, where oxygen constitutes 10% molar composition ("S$_2$ + 10% O$_2$" case); (3) a mixture of sulphur and oxygen, where oxygen constitutes 20% molar composition ("S$_2$ + 20% O$_2$" case) and (4) Ar atoms ("Ar" case). For all the cases, the MoO$_3$ film was placed in the middle of the simulation box. Apart from the MoO$_3$ film, (a) the "S$_2$" case contained 60 S$_2$ molecules, (b) the "S$_2$ + 10% O$_2$" case contained 60 S$_2$ molecules and 6 O$_2$ molecules, (c) the "S$_2$ + 20% O$_2$" case contained 60 S$_2$ molecules and 12 O$_2$ molecules, and (d) the "Ar" case contained 60 Ar atoms.

**Density functional theory (DFT) calculations**

Since we are studying gas-phase molecules, a vacuum was provided in all directions to minimize the periodic image interactions. The convergence study summarized in Tables S4 and S5 was done for the first intermediate in the pathway. Owing to the large size of the supercell in real space, a gamma point provides sufficient k-space sampling in the Brillouin zone. Based on the convergence study for the model simulating the oxygen-rich environment, a gamma point calculation, a Gaussian smearing width of 0.01 eV, and a basis set corresponding to a 500 eV kinetic energy cutoff are sufficient to reach the convergence criteria of 1 meV/atom compared to a baseline kinetic energy cutoff of 600 eV. For accurate calculations, VASP recommends 1.3× the highest ENMAX among the atomic species in the system. This corresponds to 520 eV for our system, hence explaining the use of 600 eV as the baseline cutoff. For computational efficiency, we chose to employ a 30 Å box size. The force convergence criterion for optimization steps was set to 0.01 eV/Å. Dispersion correction to the DFT energy was incorporated using Grimme's D3 method with Becke–Johnson damping[38,39]. NEB calculations were performed to determine the energy barriers between the intermediates along the minimum energy pathway (MEP)[40]. The initial guess for the MEP was generated using linear interpolation between the initial and final states. Depending on the complexity of the transition state, the number of images in the band was varied between 5 and 9. The climbing image NEB (CI-NEB) method was employed for all calculations[41], and the images were optimized until the NEB forces converged to within the tolerance criterion of 0.05 eV/Å.

## ASSOCIATED CONTENT

**Supporting Information** is available for this paper, with the following sections: Process flow and effect of changes in experimental conditions, Cut planes for computational fluid dynamics (CFD) simulations, Convergence study for the AIMD simulation parameters, Convergence study for the reaction pathway DFT calculations, Mechanism of inserting various moieties into the $Mo_xO_y$ precursor in oxygen-rich pathway towards the formation of $MoS_2$, Oxygen-rich pathway, Sulphurisation of Mo-oxysulphide ($MoO_3S_2$), Additional mechanisms explored for the oxygen-rich $SO_2$-based pathway, Optical microscopy images and CFD simulations of samples grown with variable oxygen flow-time and interval, Scanning electron microscopy (SEM) images of samples grown with variable oxygen flow-interval, Effect of change in oxygen flow-rate in O-CVD, Optical properties (Photoluminescence and Raman) of as-grown flakes, Synthesis of monolayer $MoS_2$ on sapphire at variable oxygen flow parameters.

## AUTHOR INFORMATION:

Corresponding Author: *Ananth Govind Rajan, ananthgr@iisc.ac.in, *Akshay Singh, aksy@iisc.ac.in

**Author Contributions:**

KSK and AS developed the experimental framework, while AG, MDKN, and AGR developed the theoretical framework. KSK performed the oxygen-dose dependent synthesis, with assistance from MV. BKA performed the image analysis for understanding nucleation density and flake size. KSK performed the CFD simulations. AG and MDKN carried out the AIMD and DFT simulations. KSK, AG, MDKN, AGR, and AS discussed and prepared the manuscript, with contributions from all authors.

**Data Availability**

All data created or analysed during this study are included in this paper and in the Supplementary Information. Additional supporting data are available from the corresponding author upon reasonable request.

**Acknowledgements**

KSK acknowledges discussion regarding COMSOL simulations with Abhinav Sinha. AS acknowledges funding from Indian Institute of Science (IISc) start-up grant. AS and AGR acknowledge funding from Anusandhan National Research Foundation (ANRF) grant SPR/2023/000397. KSK acknowledges DST-INSPIRE fellowship. AGR acknowledges the Infosys Foundation, Bengaluru, for an Infosys Young Investigator grant. The authors acknowledge the Micro Nano Characterization Facility (MNCF at Centre for Nano Science and Engineering (CeNSE), IISc). The authors acknowledge the financial support received from the I-STEM program funded by the Office of the Principal Scientific Adviser to the Government of India, for providing the COMSOL license.

**Table of contents**

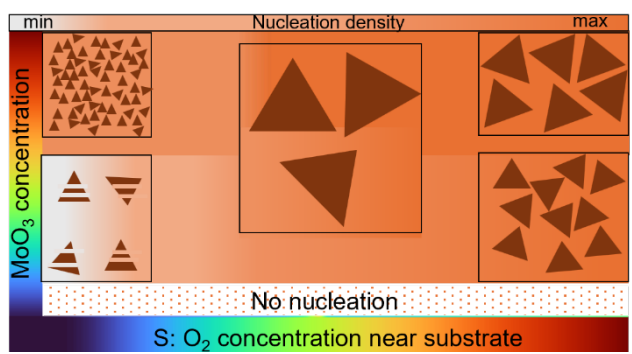

Supplementary Information

# Oxygen as a dual function regulator in MoS$_2$ CVD synthesis: enhancing precursor evaporation while modulating reaction kinetics

Keerthana S Kumar[1], Abhijit Gogoi[2#], Madhavan DK Nampoothiri[2#], Bhavesh Kumar Acharya[1], Manvi Verma[1], Ananth Govind Rajan[2*], Akshay Singh[1*]

[1]Department of Physics, Indian Institute of Science, Bengaluru, Karnataka 560012, India

[2]Department of Chemical Engineering, Indian Institute of Science, Bengaluru, Karnataka 560012, India

*Corresponding author: ananthgr@iisc.ac.in, aksy@iisc.ac.in

# These authors contributed equally.

**Table of Contents**

## 13. Synthesis of monolayer MoS$_2$ on sapphire at variable oxygen flow parameters

**S1. Process flow and effect of changes in experimental conditions**

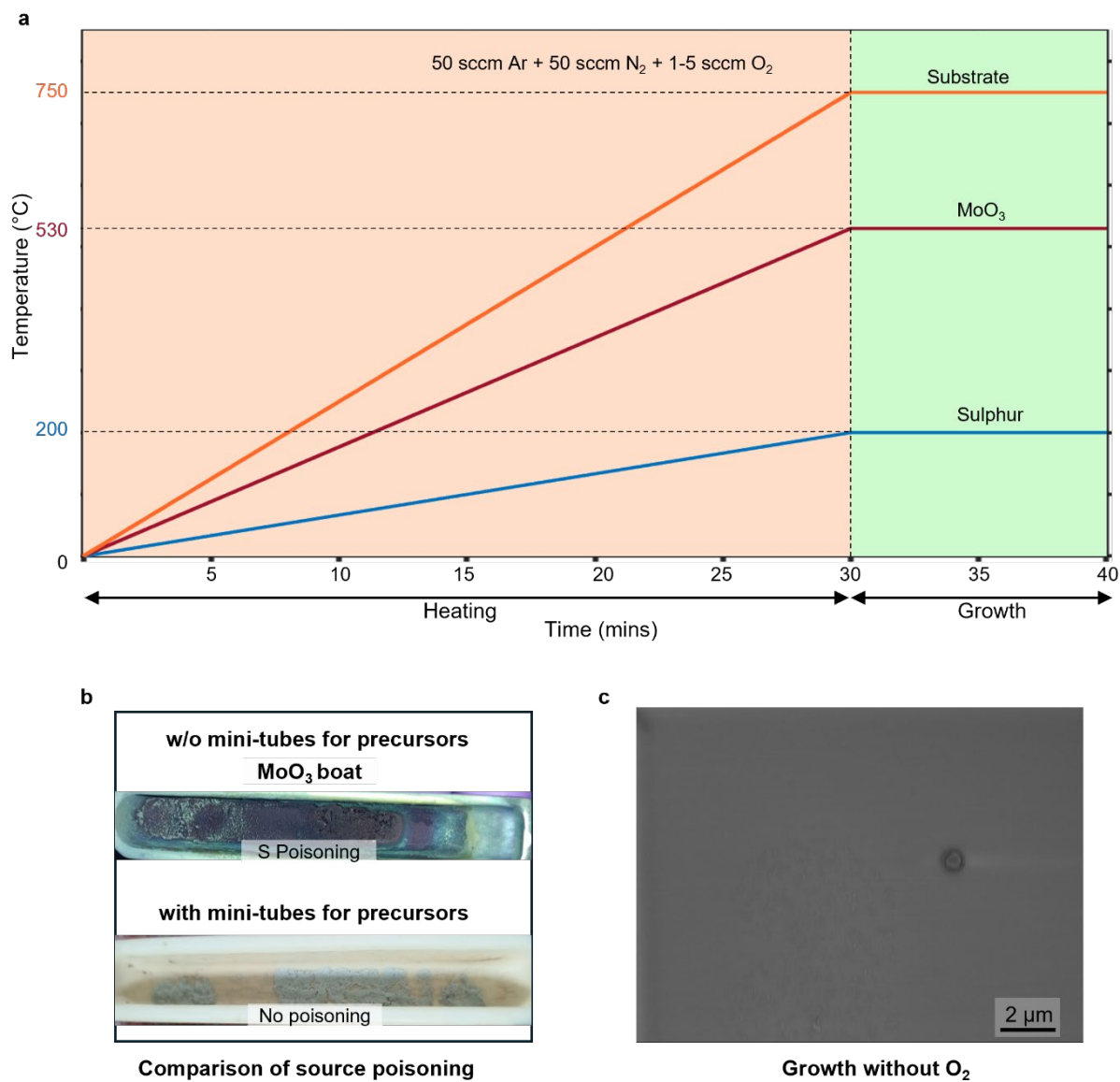

**Figure S1. Process flow and significance of experimental conditions.** (a) Process flow diagram showing heating and growth duration during O-CVD. All zones are ramped upto their respective temperatures in 30 mins, held there for 10 mins for growth and thereafter left for natural cooling. (b) Comparison of MoO$_3$ boat after growth, with and without separate mini-tubes for precursors. (c) Scanning electron microscopy (SEM) image of substrate after growth without oxygen.

## S2. Cut planes for computational fluid dynamics (CFD) simulations

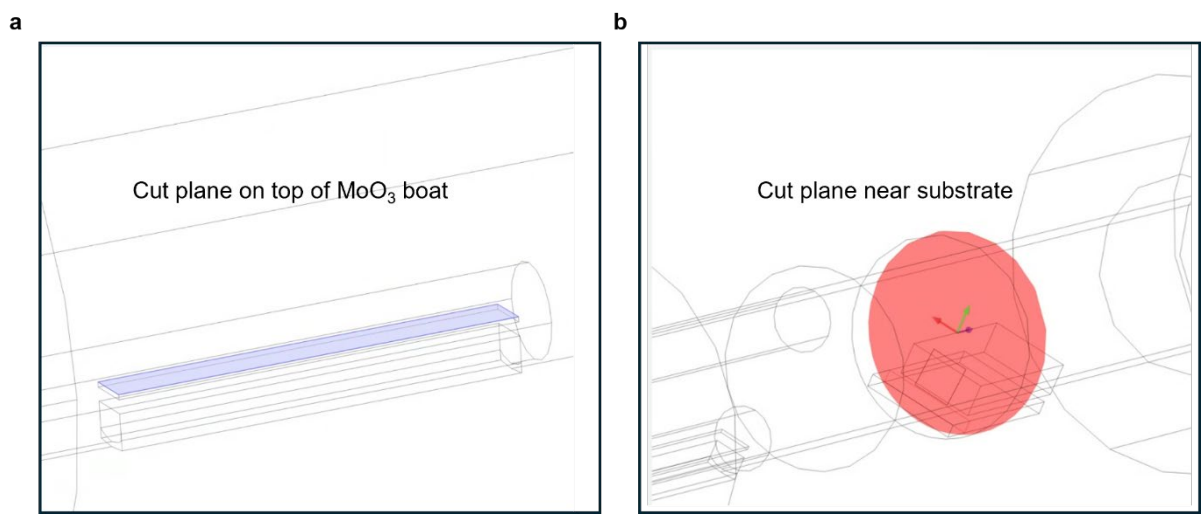

**Figure S2. Cut planes for computational fluid dynamics (CFD) simulations.** Cut planes defined in COMSOL for extracting concentration profiles of sulphur and oxygen at the (a) MoO$_3$ boat, and (b) substrate respectively.

## S3. Convergence study for the AIMD simulation parameters

| Smearing width ($\sigma$ eV) | Energy without entropy (eV) | Energy ($\sigma \to 0$ eV) | Energy difference (meV/atom) | Time (seconds) |
|---|---|---|---|---|
| 0.01 | −307.52 | −307.55 | 0 | 371.8 |
| 0.05 | −307.72 | −308.07 | 5.7 | 368.2 |
| 0.1 | −307.52 | −308.36 | 13.6 | 368.5 |
| 0.2 | −307.06 | −309.07 | 32.4 | 332.2 |

**Table S1.** Effect of the Gaussian smearing width on the DFT energy calculations. The table shows the DFT energy without electronic entropy (in eV), the energy as the Gaussian smearing width approaches zero (in eV), the energy difference per atom (in meV/atom), and the computational time (in seconds) for each smearing width (in eV).

| ENCUT (eV) | Energy difference from 600 eV value (meV/atom) | Time (seconds) |
|:---:|:---:|:---:|
| 350 | −10.8 | 227.4 |
| 400 | −6.8 | 265.3 |
| 450 | −9.8 | 328.0 |
| 500 | −6.3 | 369.0 |
| 550 | −1.7 | 445.0 |
| 600 | 0.0 | 460.0 |

**Table S2.** Basis set convergence concerning the plane-wave cutoff energy (ENCUT). The table shows the energy difference per atom relative to the ENCUT of 600 eV and the corresponding computational time.

| k-point mesh | Energy difference (meV/atom) |
|:---:|:---:|
| 1 × 1 × 1 | 0 |
| 3 × 3 × 3 | |

**Table S3.** K-point mesh convergence with respect to the 3 × 3 × 3 mesh. The table shows the energy difference per atom for the 1 × 1 × 1 mesh compared to the 3 × 3 × 3 mesh.

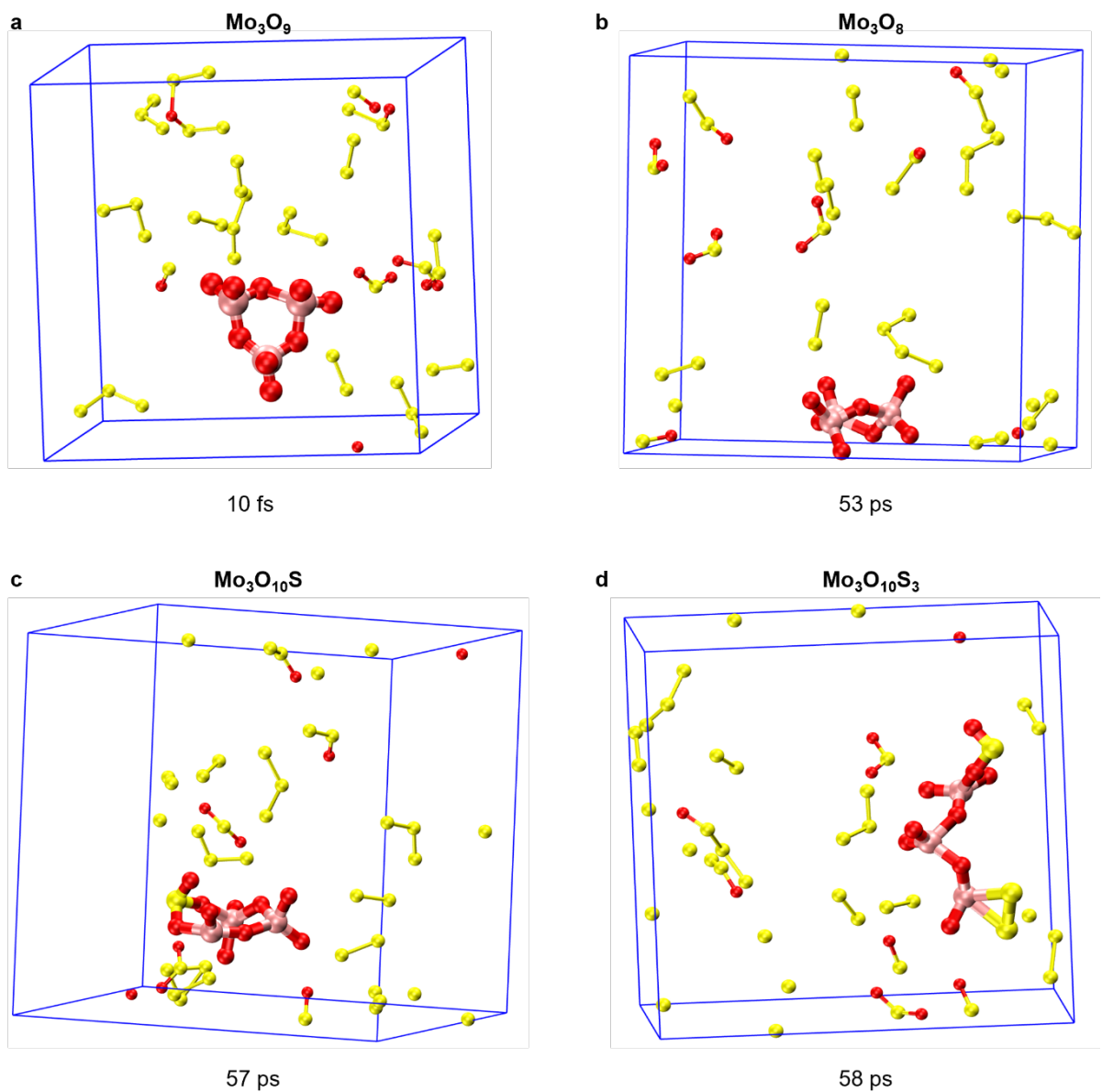

**Figure S3. Snapshots from AIMD simulation at various times.** These snapshots illustrate key intermediates in the oxygen-rich pathway outlined in Figure 3a. (a) $Mo_3O_9$ ring at 10 fs, (b) $Mo_3O_8$ formed after $S_2O$ removal at 53 ps, (c) $Mo_3O_{10}S$ chain formed after $SO_2$ insertion at 57 ps, and (d) $Mo_3O_{10}S_3$ formed after ring opening at 58 ps. Pink, red, and yellow spheres represent Mo, O, and S atoms, respectively.

## S4. Convergence study for the reaction pathway DFT calculations

| ENCUT (eV) | Energy ($\sigma \to 0$) (eV) | Energy difference (meV/atom) |
|---|---|---|
| 300 | −105.41 | 0.8 |
| 350 | −104.65 | 0.2 |
| 400 | −104.48 | 0.1 |
| 450 | −104.36 | 0 |
| 500 | −104.32 | 0 |
| 550 | −104.32 | 0 |
| 600 | −104.34 | |

**Table S4.** Basis set convergence with respect to the plane-wave cutoff energy (ENCUT). The table shows the energy as the smearing width approaches zero (in eV), the energy difference per atom relative to the respective ENCUT mentioned in the table.

| k-points | Relative energy ($\sigma \to 0$) (meV/atom) | Time (seconds) |
|---|---|---|
| $1 \times 1 \times 1$ | 0.6 | 384 |
| $2 \times 2 \times 2$ | 0 | 5028 |
| $3 \times 3 \times 3$ | | |

**Table S5.** K-point mesh convergence study. This table shows the relative energy difference with respect to the $3 \times 3 \times 3$ k-point mesh and the computational time required for different k-point meshes.

## S5. Mechanism of inserting various moieties into the Mo$_x$O$_y$ precursor in the oxygen-rich pathway towards the formation of MoS$_2$

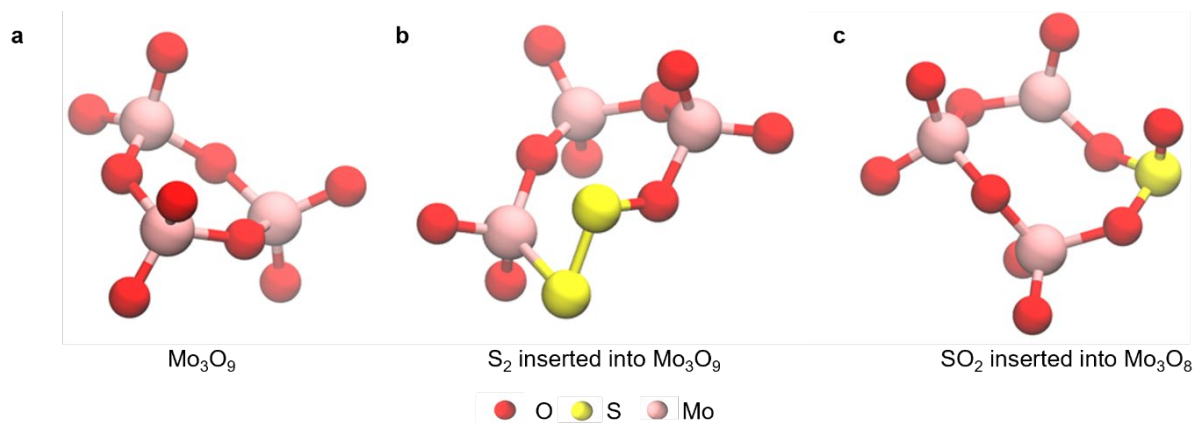

**Figure S4. Images of various moieties inserted into the ring structure.** Images of (a) Mo$_3$O$_9$, (b) S$_2$ and (c) SO$_2$ inserted into the ring structure of Mo$_3$O$_9$ and Mo$_3$O$_8$ respectively.

In our AIMD simulations for the oxygen-rich pathway, the first step involves the attachment of S$_2$ to the terminal oxygen and removing S$_2$O. This is followed by SO$_2$ insertion into the Mo$_3$O$_8$ ring (Main Text Figure 3a, Table S6). The insertion mechanism has also been observed in the oxygen-free pathway (without oxygen assistance) by Lei et al.[1], where S$_2$ gets inserted into the Mo$_3$O$_9$ ring in the first step. Figure S4 represents the previously described insertion mechanism. Further details are explained in the upcoming subsections.

## S6. Oxygen-rich pathway

| Sl. no. | Reaction step | Chemical formula | $\Delta E$ (eV) | $E_a$ (eV) |
|---|---|---|---|---|
| 1 | S$_2$ addition | Mo$_3$O$_9$ + S$_2$ ⟶ Mo$_3$O$_9$S$_2$ | −0.03 | no barrier |
| 2 | S$_2$O removal | Mo$_3$O$_9$S$_2$ ⟶ Mo$_3$O$_8$ + S$_2$O | 1.79 | 1.79 |
| 3 | SO$_2$ insertion | Mo$_3$O$_8$ + SO$_2$ ⟶ Mo$_3$O$_{10}$S | −0.11 | 0.15 |
| 4 | S$_2$ addition | Mo$_3$O$_{10}$S + S$_2$ ⟶ Mo$_3$O$_{10}$S$_3$ | −3.15 | no barrier |
| 5 | Ring opening | Mo$_3$O$_{10}$S$_3$ ⟶ Mo$_3$O$_{10}$S$_3$ | 0.78 | 0.78 |

| 6 | $S_2$ insertion into chain | $Mo_3O_{10}S_3 + S_2 \longrightarrow Mo_3O_{10}S_5$ | −0.26 | 0.92 |
|---|---|---|---|---|
| 7 | Chain breaking | $Mo_3O_{10}S_5 \longrightarrow Mo_2O_5S_4 + MoO_5S$ | 1.25 | 1.72 |
| 8 | $S_2$ addition and $SO_2$ removal | $MoO_5S + S_2 \longrightarrow MoO_3S_2 + SO_2$ | −0.09 | 0.57 |

**Table S6.** Energetics and kinetics of the oxygen-rich pathway, showing the energy changes ($\Delta E$) and activation energies ($E_a$) for each reaction step up to chain breaking, as outlined in Main Text Figure 3a.

As discussed in the oxygen-rich pathway, the process begins with the addition of a disulphur molecule ($S_2$) to the Mo-based ring structure. This step is slightly exothermic, with minimal energy release. Note that this contrasts with the oxygen-free pathway reported in Ref[1], wherein the addition of disulphur to the Mo ring was highly exothermic. In this reference, the sulphur atoms are pointing inward to the ring, plausibly explaining its lower energy. The removal of oxygen creates enough space to accommodate the larger $SO_2$ molecule within the ring ($Mo_3O_8$), allowing for more favourable insertion. This is also evident from the fact that the $SO_2$ molecule insertion into the $Mo_3O_8$ ring is slightly exothermic with a low barrier for transformation. Subsequently, $S_2$ addition into the ring occurs, releasing a significant amount of energy, followed by a ring opening step, which is endothermic due to the bond breaking required to open the ring. Following the ring opening, the sulphurisation proceeds with the insertion of $S_2$ into the chain, which is mildly exothermic. The chain further breaks into double-Mo-containing and single-Mo-containing chains, which is an endothermic process with a high energy barrier and is the rate-determining step in the oxygen-rich pathway (Main Text Figure 3a, Table S6).

## S7. Sulphurisation of Mo-oxysulphide (MoO$_3$S$_2$)

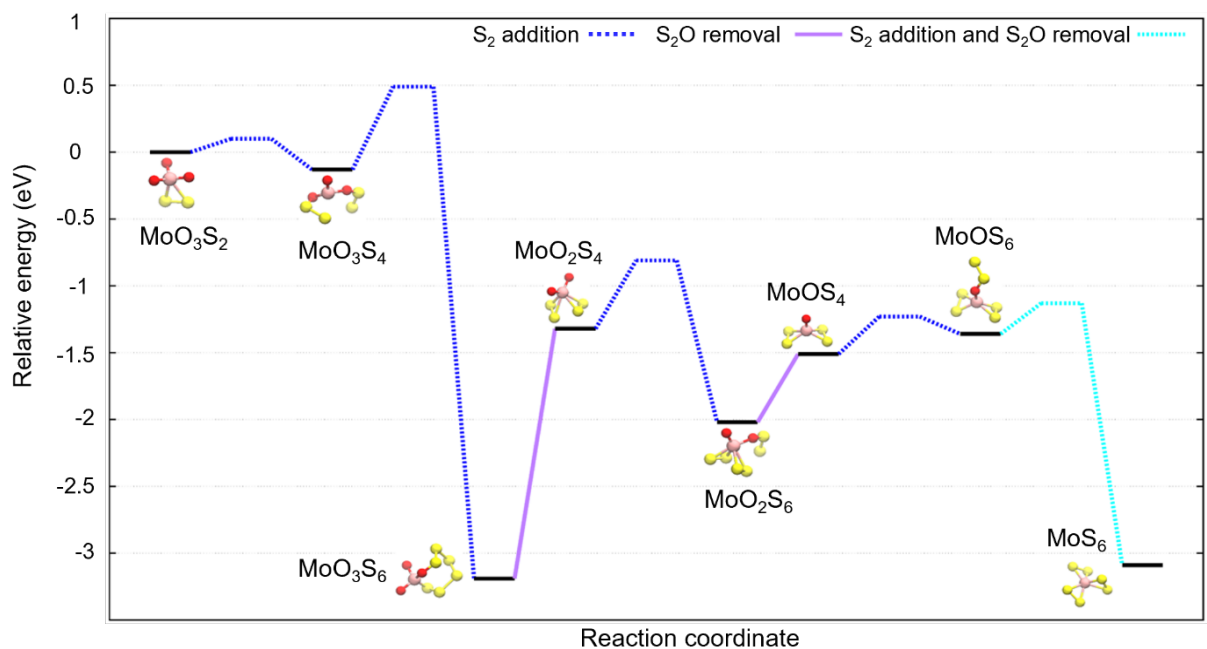

**Figure S5. Reaction pathway diagram for the sulphurisation of MoO$_3$S$_2$ to finally form MoS$_6$.** Key steps include S$_2$ addition and S$_2$O removal. The pathway begins with MoO$_3$S$_2$, the final intermediate after the chain-breaking step in the oxygen-rich pathway, and progresses to MoS$_6$, the precursor for MoS$_2$ formation as reported in literature[1]. Dashed lines indicate reactions with transition states, while bold lines represent reactions without transition states, showing uphill or downhill processes. Horizontal bold black lines denote intermediate species, with adjacent smaller images displaying the corresponding molecular structures. Pink, red, and yellow spheres represent Mo, O, and S atoms, respectively. Table S7 summarizes the energetics and activation barriers for these reactions.

| Sl. no. | Reaction step | Chemical formula | ΔE (eV) | E$_a$ (eV) |
|---|---|---|---|---|
| 1 | S$_2$ addition | MoO$_3$S$_2$ + S$_2$ ⟶ MoO$_3$S$_4$ | −0.13 | 0.10 |
| 2 | S$_2$ addition | MoO$_3$S$_4$ + S$_2$ ⟶ MoO$_3$S$_6$ | −3.06 | 0.62 |
| 3 | S$_2$O removal | MoO$_3$S$_6$ ⟶ MoO$_2$S$_4$ + S$_2$O | 1.87 | 1.87 |
| 4 | S$_2$ addition | MoO$_2$S$_4$ + S$_2$ ⟶ MoO$_2$S$_6$ | −0.70 | 0.51 |

| 5 | S$_2$O removal | MoO$_2$S$_6$ ⟶ MoOS$_4$ + S$_2$O | 0.51 | 0.51 |
| 6 | S$_2$ addition | MoOS$_4$ + S$_2$ ⟶ MoOS$_6$ | 0.15 | 0.28 |
| 7 | S$_2$ addition | MoOS$_6$ + S$_2$ ⟶ MoS$_6$ + S$_2$O | −1.73 | 0.23 |

**Table S7.** Energetics and kinetics of the sulphurization pathway of the single-Mo-containing chain, showing the energy changes ($\Delta E$) and activation energies ($E_a$) for each reaction step up to chain breaking, as outlined in Figure S5.

## S8. Additional mechanisms explored for the oxygen-rich SO$_2$-based pathway

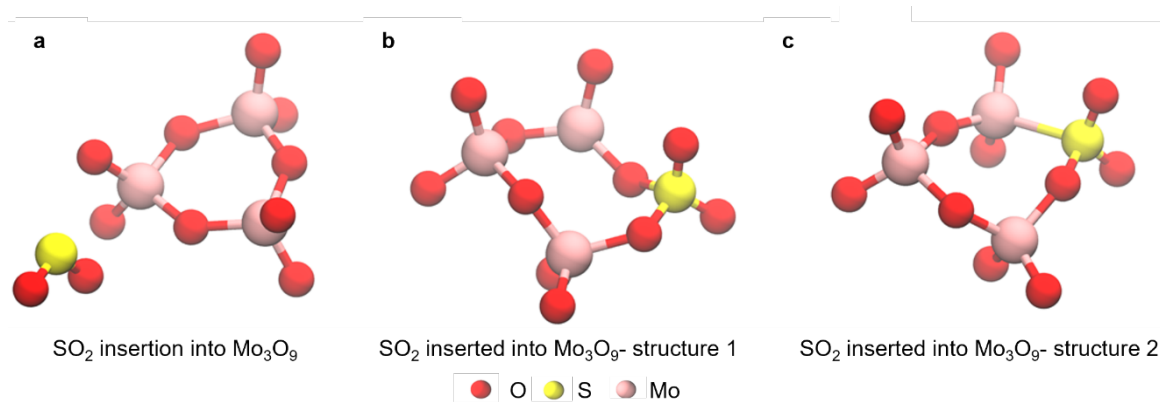

**Figure S6. Insertion of SO$_2$ into Mo$_3$O$_9$ ring structure.** (a) SO$_2$ insertion into Mo$_3$O$_9$. (b) Structure formed with SO$_2$ positioned between two oxygen atoms in the Mo$_3$O$_9$ ring. (c) Structure formed with SO$_2$ positioned between a molybdenum and oxygen atom in the Mo$_3$O$_9$ ring.

The reaction shown in Figure S6 attempts to insert SO$_2$ directly into the Mo$_3$O$_9$ ring. This is in contrast to the mechanism observed in the AIMD simulations, where SO$_2$ is inserted into a Mo$_3$O$_8$ ring. The Mo$_3$O$_8$ ring is formed after an initial S$_2$ insertion in the Mo$_3$O$_9$ ring, followed by the removal of an S$_2$O moiety.

The activation barrier for SO$_2$ insertion into the Mo$_3$O$_8$ ring is calculated to be 0.15 eV, with an associated energy difference of −0.11 eV, as shown in Table S6. In contrast, the energy difference for the direct insertion of SO$_2$ into the Mo$_3$O$_9$ ring in between oxygen atoms is much higher, calculated to be 2.16 eV. The energy difference for the insertion of SO$_2$ in between molybdenum and oxygen atoms in $Mo_3O_9$ is found to be 1.08 eV. This indicates that the direct insertion of $SO_2$ into the $Mo_3O_9$ ring is energetically unfavorable. The difficulty in inserting $SO_2$ into the $Mo_3O_9$ ring arises because $SO_2$ is larger than $S_2$, resulting in strain within the ring upon insertion. Successful insertion can occur if an oxygen atom is removed, i.e., by converting the $Mo_3O_9$ ring into a $Mo_3O_8$ ring, as discussed in section S7.

Similarly, reactions 4, 6 and 7 were also optimized with $S_2$ due to high energy barriers during reactions with $SO_2$ as given in Table S8, S9 and Main Text Figure 3b, c.

| Sl. no. | Reaction step | Chemical formula | $\Delta E$ (eV) | $E_a$ (eV) |
|---|---|---|---|---|
| a | $S_2$ addition | $Mo_3O_{10}S + S_2 \longrightarrow Mo_3O_{10}S_3$ | −3.15 | no barrier |
| b | $SO_2$ addition | $Mo_3O_{10}S + SO_2 \longrightarrow Mo_3O_{12}S_2$ | 0.26 | 0.72 |

**Table S8.** Energetics and kinetics of the modified elementary reaction 4 in the oxygen-rich pathway outlined in Main Text Figure 3b.

| Sl. no. | Reaction step | Chemical formula | $\Delta E$ (eV) | $E_a$ (eV) |
|---|---|---|---|---|
| a1 | $S_2$ insertion into chain | $Mo_3O_{10}S_3 + S_2 \longrightarrow Mo_3O_{10}S_5$ | −0.26 | 0.92 |
| a2 | Chain breaking after $S_2$ insertion | $Mo_3O_{10}S_5 \longrightarrow Mo_2O_5S_4 + MoO_5S$ | 1.25 | 1.72 |
| b1 | $SO_2$ insertion into chain | $Mo_3O_{10}S_3 + SO_2 \longrightarrow Mo_3O_{12}S_4$ | 0 | 0.33 |
| b2 | Chain breaking after $SO_2$ insertion | $Mo_3O_{12}S_4 \longrightarrow Mo_2O_7S_3 + MoO_5S$ | 2.59 | 2.89 |

**Table S9.** Energetics and kinetics of the modified elementary reactions 6 and 7 in the oxygen-rich pathway outlined in Main Text Figure 3c.

**S9. Optical microscopy (OM) images and CFD simulations of samples grown with variable oxygen flow-time and interval**

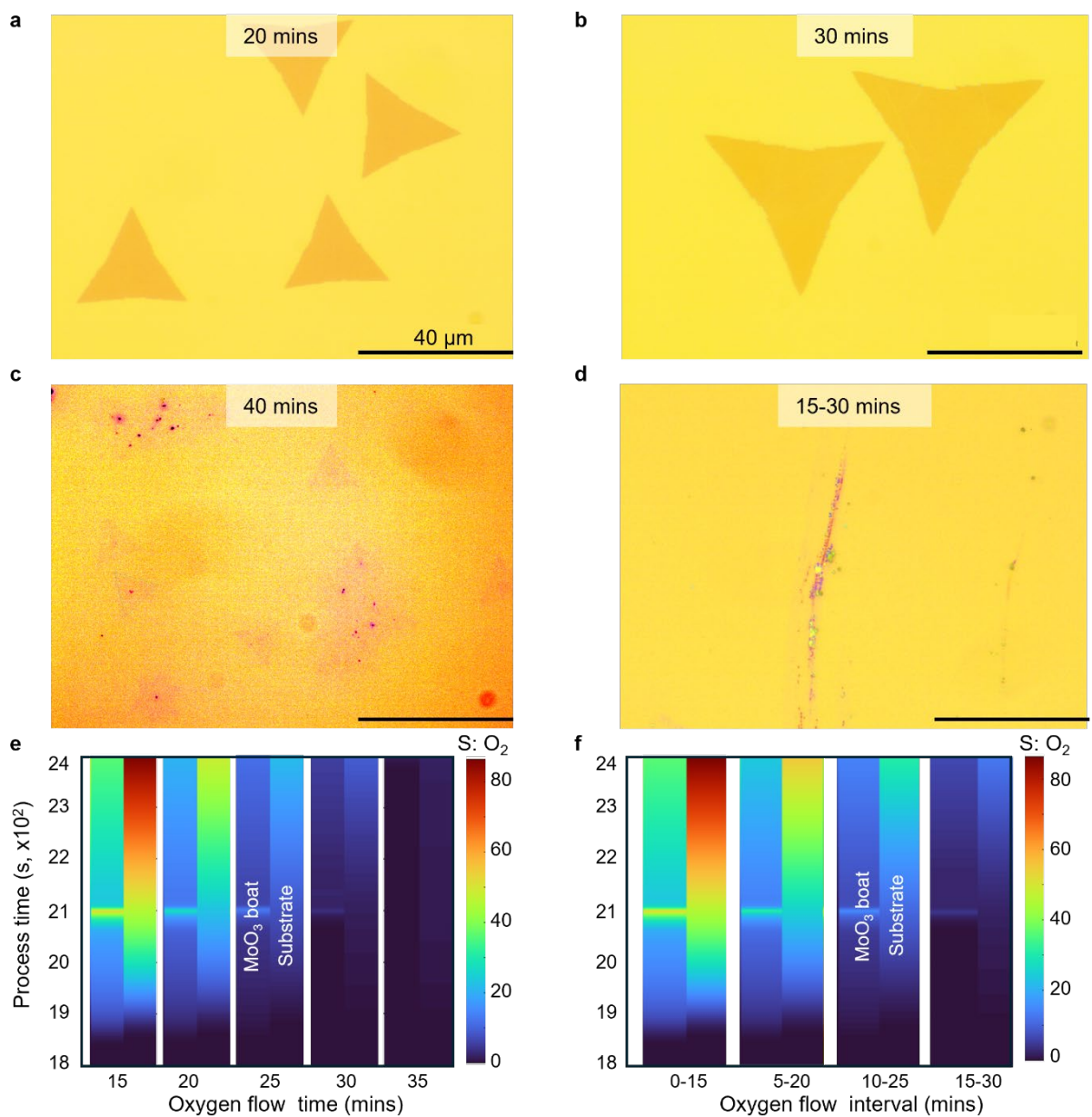

**Figure S7. Effect of change in oxygen flow-time and flow-interval during O-CVD.** Optical microscopy (OM) images of the substrate after growth with 1 sccm oxygen flow-rate and (a)-(c) oxygen flow-time of 20 mins, 30 mins and 40 mins respectively, and (d) oxygen flow-interval between 15-30 mins. Variation of sulphur-to-oxygen (S:$O_2$) concentration ratio at $MoO_3$ boat and substrate with 1 sccm oxygen flow-rate and variable oxygen (e) flow-time and (f) flow-interval. Note that growth starts at 1800 s (30 mins). CFD simulations were performed to calculate S:$O_2$ concentration ratios. Note that the range for S:$O_2$ ratio for (e) and (f) is 0-80.

## S10. Scanning electron microscopy (SEM) images of samples grown with variable oxygen flow-interval

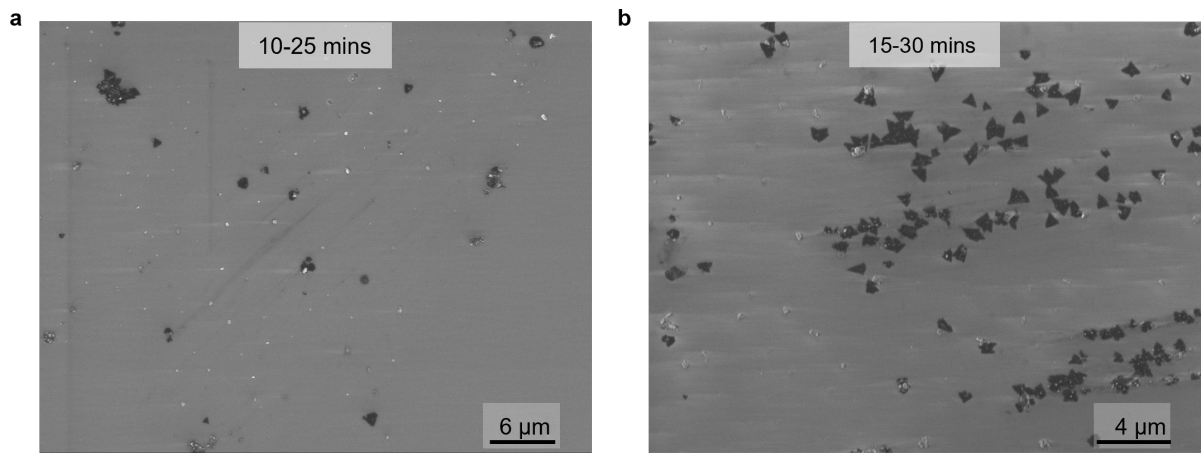

**Figure S8. SEM images of samples synthesized with variable oxygen flow-interval.** SEM images of substrate after growth with 1 sccm oxygen flow-rate and oxygen flow during (a) 10-25 mins, and (b) 15-30 mins respectively. Total growth time was fixed at 40 mins.

The growth was nearly invisible (using OM images) for samples synthesized with a flow-interval of 10-25 mins and 15-30 mins (Figure 4b, S7d). However as can be seen from Figure S8a,b there is sparse growth in these samples.

## S11. Effect of change in oxygen flow-rate in O-CVD

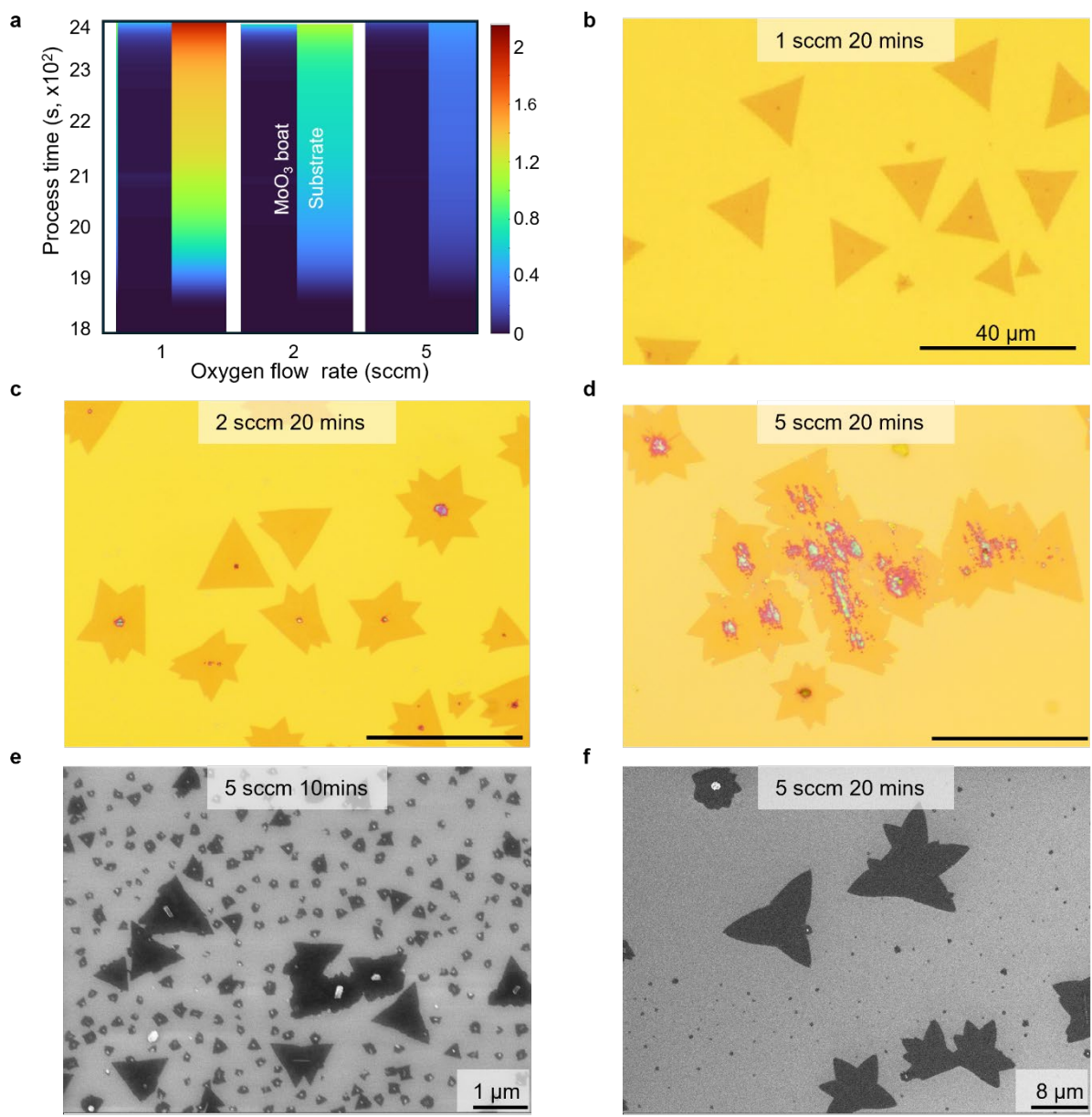

**Figure S9. Effect of change in oxygen flow-rate in O-CVD.** (a) Variation of S:O$_2$ ratio at MoO$_3$ boat and substrate with variable oxygen flow-rates. OM images of substrate after growth with oxygen flow-rate of (b) 1 sccm, (c) 2 sccm, and (d) 5 sccm respectively with 20 mins growth time and 0-35 mins oxygen flow-time. SEM images of substrate after growth with 5 sccm oxygen flow-rate, 0-35 mins oxygen flow-time and growth time of (e) 10 mins and (f) 20 mins respectively.

It can be observed from Figure S9e that the dot like structures seen in OM images (Figure 4c) are also monolayer flakes with varying shapes. At an increased growth time, some of the flakes grew bigger with multiple edges (Figure S9f) as discussed in the Main Text.

**S12. Optical properties (Photoluminescence and Raman) of as-grown flakes**

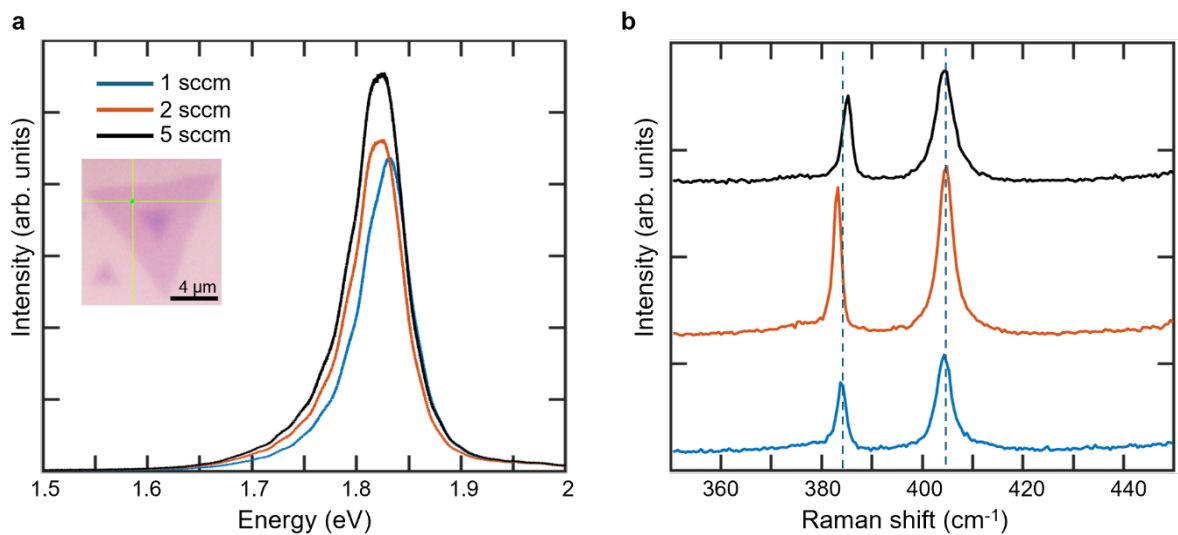

**Figure S10. Optical properties of MoS$_2$ synthesized at different oxygen flow-rates.** (a) Photoluminescence spectra and (b) Raman spectra of samples synthesized at different oxygen flow-rates, fixed 20 mins growth time, and 35 mins oxygen-flow time.

## S13. Synthesis of monolayer MoS₂ on sapphire at variable oxygen flow parameters

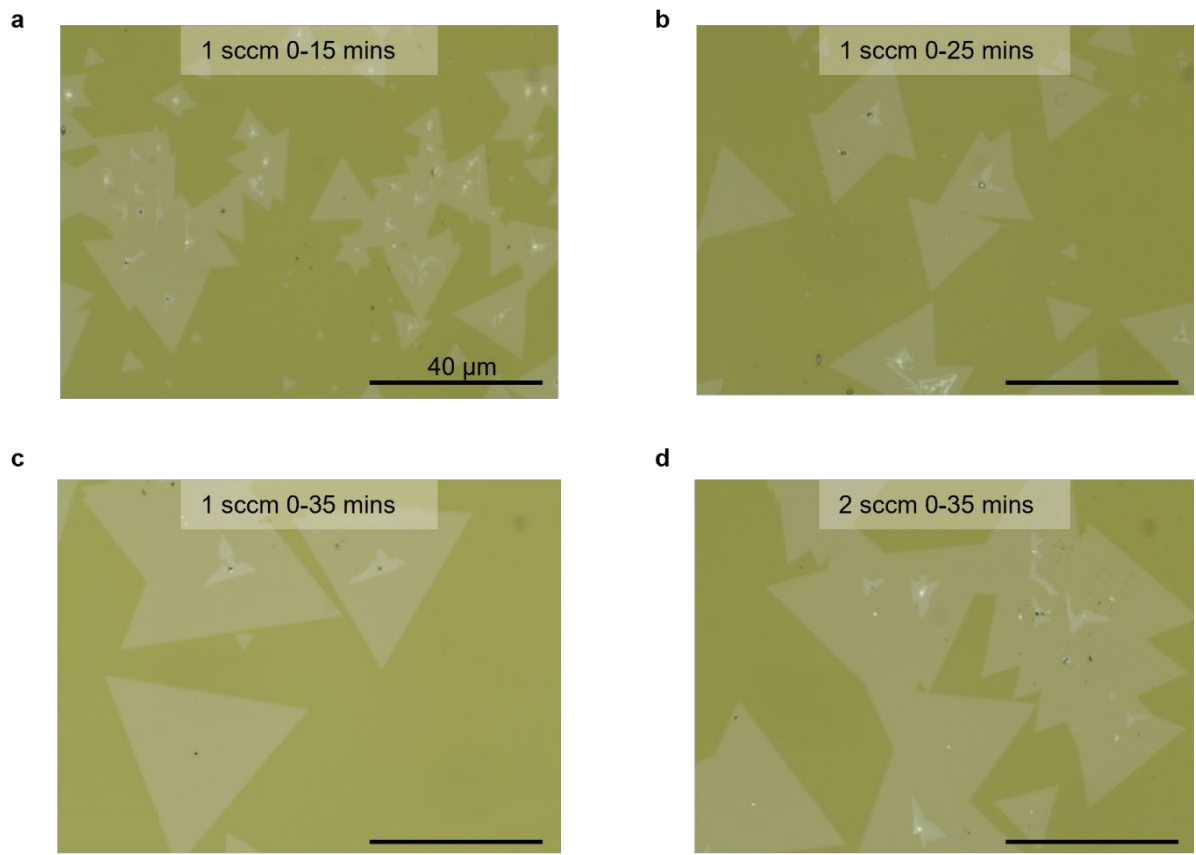

**Figure S11. Optical microscopy images of MoS₂ synthesized on sapphire substrate at different oxygen flow parameters.** OM images of sapphire substrates after growth at 1 sccm oxygen flow-rate and flow-time of (a) 0-15 mins, (b) 0-25 mins and (c) 0-35 mins. (d) OM image of sapphire substrate after growth at 2 sccm oxygen flow-rate and 0-35 mins flow-time. Growth time was fixed at 20 mins.